\documentclass{jetpl}
\twocolumn

\lat

\usepackage[english]{babel}
\usepackage{pgfplots,pgfplotstable}
\usepackage{caption}
\addto\captionsenglish{\def\refname{}}
\usepackage{cite}

\def\dd{{\mathrm d}}
\newcommand{\SANC}{\texttt{SANC}{}}
\newcommand{\MCSANC}{\texttt{MCSANC}}

\newcommand{\ReneSANCe}{\texttt{ReneSANCe}}

\newcommand{\sss}[1]{\scriptscriptstyle{#1}}
\newcommand{\ds}{\displaystyle}

\title{EW one-loop corrections to the longitudinally polarized Drell--Yan scattering. (I). The Neutral current case}
\rtitle{EW one-loop corrections to the longitudinally polarized Drell--Yan scattering. (I). The Neutral current case}
\sodtitle{EW one-loop corrections to the longitudinally polarized Drell--Yan scattering. (I). The Neutral current case}

\author{S.\,Bondarenko$^1$, Ya.\,Dydyshka$^{2,3}$, L.\,Kalinovskaya$^2$, R.\,Sadykov$^2$, V.\,Yermolchyk$^{2,3}$}
\rauthor{S.\,Bondarenko, Ya.\,Dydyshka, L.\,Kalinovskaya, R.\,Sadykov, V.\,Yermolchyk}
\sodauthor{S.\,Bondarenko, Ya.\,Dydyshka, L.\,Kalinovskaya, R.\,Sadykov, V.\,Yermolchyk}

\address{\small$^1$ Bogoliubov Laboratory of Theoretical Physics, JINR, 
                 141980 Dubna, Moscow region, Russia}
\address{\small$^2$ Dzhelepov Laboratory of Nuclear Problems, JINR,  
                 141980 Dubna, Moscow region, Russia}
\address{\small$^3$ Institute for Nuclear Problems, Belarusian State University, Minsk, 220006  Belarus}

\date{\today}

\abstract{
Complete one-loop electroweak corrections 
to neutral current Drell-Yan  process $p p \to \ell^+\ell^- X$ 
are presented for the case of longitudinal polarization of initial
particles. 
Cross sections for longitudinally
polarized protons allow us to estimate
different combinations of polarized quark distributions from single- and double-spin asymmetries. 
Numerical impact of electroweak next-order corrections
to asymmetries as function
of the vector boson rapidity 
and lepton pseudorapidities
in the hadron-hadron centre-of-mass frame
using the MC generator ReneSANCe
is thoroughly studied.
}

\begin{document}
\maketitle

\section{Introduction}
Theoretical calculations of
one-loop QED and electroweak (EW) radiative corrections (RC)
for Drell-Yan (DY) \cite{Drell:1970wh} processes at high energy hadronic colliders were performed by several groups,
see papers 
~\cite{Mosolov:1981xk,
Soroko:1990ug,
Wackeroth:1996hz,
Baur:1998kt,
Dittmaier:2001ay,
Baur:2001ze,
Baur:2002fn,
Baur:2004ig,
CarloniCalame:2006zq} and references therein.

The measurement of the
DY cross section in polarized hadron-hadron collisions
would provide important information about the polarization of the quark sea in the nucleon which is currently analyzed only 
from the deep inelastic scattering data experiments, such as  
$l-p$
scattering at HERA, SMC spin-muon collaboratios at CERN and etc.
Computer codes relevant for the description of polarized processes for these experiments were created in our group, namely,
the $\mu$ela code~\cite{Bardin:1997nc}
for investigation of the Spin dependent structure function $g_1(x)$ of the deuteron from polarized deep inelastic muon scattering ~\cite{SpinMuonSMC:1997ixm},
and the {\tt polHECTOR} code~\cite{Bardin:1996ch}
for deep inelastic scattering 
with longitudinally and transversely polarized nucleon  
for the HERMES experiment in HERA.
The weak corrections were small and neglected.
 
The research of longitudinally polarized proton-proton collisions at the QCD level has been carried out  
in several papers.
The most important works for the longitudinally polarized DY process at 
QCD level are:
complete analytical results for mass differential Drell-Yan type cross-sections
\cite{Kamal:1995as,Kamal:1997fg},
investigation of the lepton helicity distributions 
\cite{Kodaira:2003tq,Bourrely:1994sc},
 complete calculations of the 
 ${\cal O}(\alpha_s)$ corrections 
 in the MS–scheme
\cite{Gehrmann:1997pi},
study of double and single spin asymmetries
\cite{Gehrmann:1997ez}.

This article is the next step in the series of papers devoted to DY processes in $pp$ mode in Monte Carlo (MC) generator {\ReneSANCe}~\cite{Bondarenko:2022mbi} and integrator {\MCSANC}~\cite{Bardin:2012jk,Bondarenko:2013nu,Arbuzov:2015yja}.
 In the last one, we presented description of 
 implementation  DY
 processes to simulate processes at hadron-hadron colliders with allowance for electroweak (EW) and QCD
corrections with the next-to-leading order (NLO) accuracy and also higher-order EW corrections through 
$\Delta \rho$ parameter.

In this paper we show results 
of NLO EW corrections
for the neutral current (NC) massive lepton pair production
in longitudinally polarized proton-proton collisions
obtained by the MC event generator {\ReneSANCe}:
\begin{equation}
\label{DYNC}
pp \to Z X \to \ell^+\ell^-X.
\end{equation}

{\SANC} team has the advantage of experience in calculation of one-loop EW corrections using helicity approach
\cite{Andonov:2004hi}.
This makes it quite easy to implement
 calculations of the polarized effects.
The calculations are based on the {\SANC} (Support for Analytic and Numeric Calculations
for experiments at colliders) modules for DY NC processes
\cite{Arbuzov:2007db}.

We study the sensitivity of the single- and double-spin asymmetries onto the magnitude and behaviour of NLO EW corrections in case of longitudinal polarization.

The paper is organized as follows: in Section 2  we define the
observables for polarized DY process,
numerical results are presented in Section 3
and 
finally, Section 4 contains the conclusion.

\section{Differential cross section}

The differential cross section of the DY process at the hadronic level 
can be obtained from convolution of the partonic cross section
with quark density functions:
\begin{eqnarray}
\label{sigpp}
&& \frac{\dd\sigma^{pp\to l\bar{l} X}(s,c)}{\dd c} 
= \sum\limits_{q_1q_2}\int\limits_{0}^{1} \int\limits_{0}^{1} 
\dd x_1\; \dd x_2\; \bar{q}_1(x_1,M^2) 
\nonumber \\ && \quad \times
\bar{q}_2(x_2,M^2)
\frac{\dd\hat{\sigma}^{q_1\bar{q}_2\to l\bar{l}}(\hat{s},\hat{c})}
{\dd\hat{c}}{\mathcal J}\Theta(c,x_1,x_2),
\end{eqnarray}
where the step function $\Theta(c,x_1,x_2)$ defines the phase space domain 
corresponding to the given event selection procedure.

At one-loop level the partonic differential
cross-section can be written as follows:
\begin{eqnarray}
\hat{\sigma}^{\text{1-loop}} & = & \hat{\sigma}^{\mathrm{Born}} + \hat{\sigma}^{\mathrm{virt}}(\lambda) + \hat{\sigma}^{\mathrm{soft}}(\lambda, \omega) \nonumber \\&&
+ \hat{\sigma}^{\mathrm{hard}}(\omega) + \hat{\sigma}^{\mathrm{Subt}}, 
\label{loopxsec}
\end{eqnarray}
where one due to the contribution of the Born level cross section $\hat{\sigma}^{\mathrm{Born}}$,
one due to $\hat{\sigma}^{\mathrm{virt}}$ virtual (loop) corrections,
one due to soft photon $\hat{\sigma}^{\mathrm{soft}}$ emission,
and one due to  $\hat{\sigma}^{\mathrm{hard}}$ the hard photon emission part (with energy $E_{\gamma} > \omega$)
(with the aid of the soft–hard separator - $\omega$ and the fuxiliary parameters $\lambda$ (fictitious "photon mass" which regularizes infrared divergences).
The special term  $\hat{\sigma}^{\mathrm{Subt}}$ stands for subtraction of collinear quark mass singularities.
To perform the subtraction procedure at the partonic
level cross section
we proceed in the same way as in our papers ~\cite{Arbuzov:2005dd, Arbuzov:2007db}.
The partonic cross section is taken in the center-of-mass reference 
frame of initial quarks/antiquarks, where the cosine of the muon scattering angle $\hat{c}$ is defined. 

We estimate following sets of fully polarized 
$\sigma^{++},\sigma^{+-},\sigma^{-+},\sigma^{--}$ components of the 
hadron-hadron 
cross section:
\begin{eqnarray}
\sigma &=&
\frac{1}{4}\left(\sigma^{++}+\sigma^{+-}+\sigma^{-+}+\sigma^{--}\right),\\
\Delta\sigma_{\mathrm {L}} &=&
\frac{1}{4}\left(\sigma^{++}+\sigma^{+-}-\sigma^{-+}-\sigma^{--}\right),\\
\Delta\sigma_{\mathrm {LL}} &=&
\frac{1}{4}\left(\sigma^{++}-\sigma^{+-}-\sigma^{-+}+\sigma^{--}\right),
\end{eqnarray}
where $\sigma=\sigma^{00}$ is unpolarized one.
These asymmetries appear if at least one of incoming hadrons is polarized.

We use following definitions  for {\it the single-spin asymmetry}
\begin{eqnarray}
{A}_{\mathrm{L}}(I)
 = \frac{{\ds \Delta d\sigma_{\mathrm {L}}}/
    {\ds dI}}
    {{\ds d\sigma}/
    {\ds dI}},
\end{eqnarray}
and for {\em the double-spin asymmetry}  
\begin{eqnarray}
{A}_{\mathrm{LL}}(I)
 = \frac{{\ds \Delta d\sigma_{\mathrm {LL}}}/
    {\ds dI}}
    {{\ds d\sigma}/
    {\ds dI}}.
\end{eqnarray}
Variable $I$ is the $Z$ boson rapidity
\begin{eqnarray}
{\mathrm y}_{\mathrm \sss Z} =
\frac{1}{2}
\ln\frac{\ds E_{\ell^+\ell^-} + p^z_{\ell^+\ell^-}}
        {\ds E_{\ell^+\ell^-} - p^z_{\ell^+\ell^-}},
\end{eqnarray}
($E_{\ell^+\ell^-}$ and $p^z_{\ell^+\ell^-}$ are the energy and $z$-component of a momentum of the $\ell^+\ell^-$ pair
in the laboratory frame)
or the lepton pseudorapidity 
\begin{eqnarray}
\eta_{\ell^\pm} =
 -\ln \tan\frac{\ds \vartheta_{\ell^\pm}}{\ds 2}.
\end{eqnarray}
Here $\vartheta_{\ell^\pm}$ is the angle of the $\ell^\pm$
in the laboratory frame.

\section{Numerical results}
\label{sec:validation}

\subsection{Input parameters}

Numerical calculations were performed in the $\alpha(0)$ schemes and the following set of input parameters was used:
\begin{eqnarray}
&&\alpha^{-1}(0)= 137.035999084, \nonumber \\
&&G_F = 1.1663787\times10^{-5} \; \text{GeV}^{-2}, \nonumber\\
&&M_W = 80.379 \; \text{GeV}, ~ M_Z = 91.1876 \; \text{GeV}, \nonumber\\
&&M_H = 125.25 \; \text{GeV}, \nonumber\\
&&\Gamma_W = 2.085 \; \text{GeV}, ~ \Gamma_Z = 2.4952 \; \text{GeV}, \nonumber\\
&&|V_{ud}| = 0.9737, ~ |V_{us}| = 0.2252, \nonumber\\
&&|V_{cd}| = 0.221, ~ |V_{cs}| = 0.987,
~ |V_{cb}| = 0, ~ |V_{ub}| = 0, \nonumber\\
&&m_e = 0.51099895 \; \text{MeV}, ~ m_\mu = 0.1056583745 \; \text{GeV}, \nonumber\\
&&m_\tau = 1.77686 \; \text{GeV},\nonumber\\
&&m_d = 0.066 \; \text{GeV}, ~ m_u = 0.066 \; \text{GeV},\nonumber\\
&&m_s = 0.15 \; \text{GeV}, ~ m_c = 1.67 \; \text{GeV},\nonumber\\
&&m_b = 4.78 \; \text{GeV}, ~ m_t = 172.76 \; \text{GeV}.
\end{eqnarray}
The values of the parameters were taken from PDG-2020~\cite{ParticleDataGroup:2020ssz}, except for the masses of light quarks (u, d and s) which were chosen as in~\cite{Dittmaier:2009cr}.

Following cuts were also applied ($\ell = e, \; \mu$):
\begin{eqnarray}
\label{Cuts}
pp \to \ell^+\ell^-X: && \quad p_\perp(\ell^\pm) > 25 \; \text{GeV}, \quad |\eta(\ell^\pm)| < 2.5, \nonumber \\ &&  M(\ell^+\ell^-) > 50 \; \text{GeV}. \nonumber
\end{eqnarray}

\subsection{Differential distributions}

\begin{figure*}[!ht]
\begin{tabular}{cc}
    \includegraphics[width=0.51\textwidth]{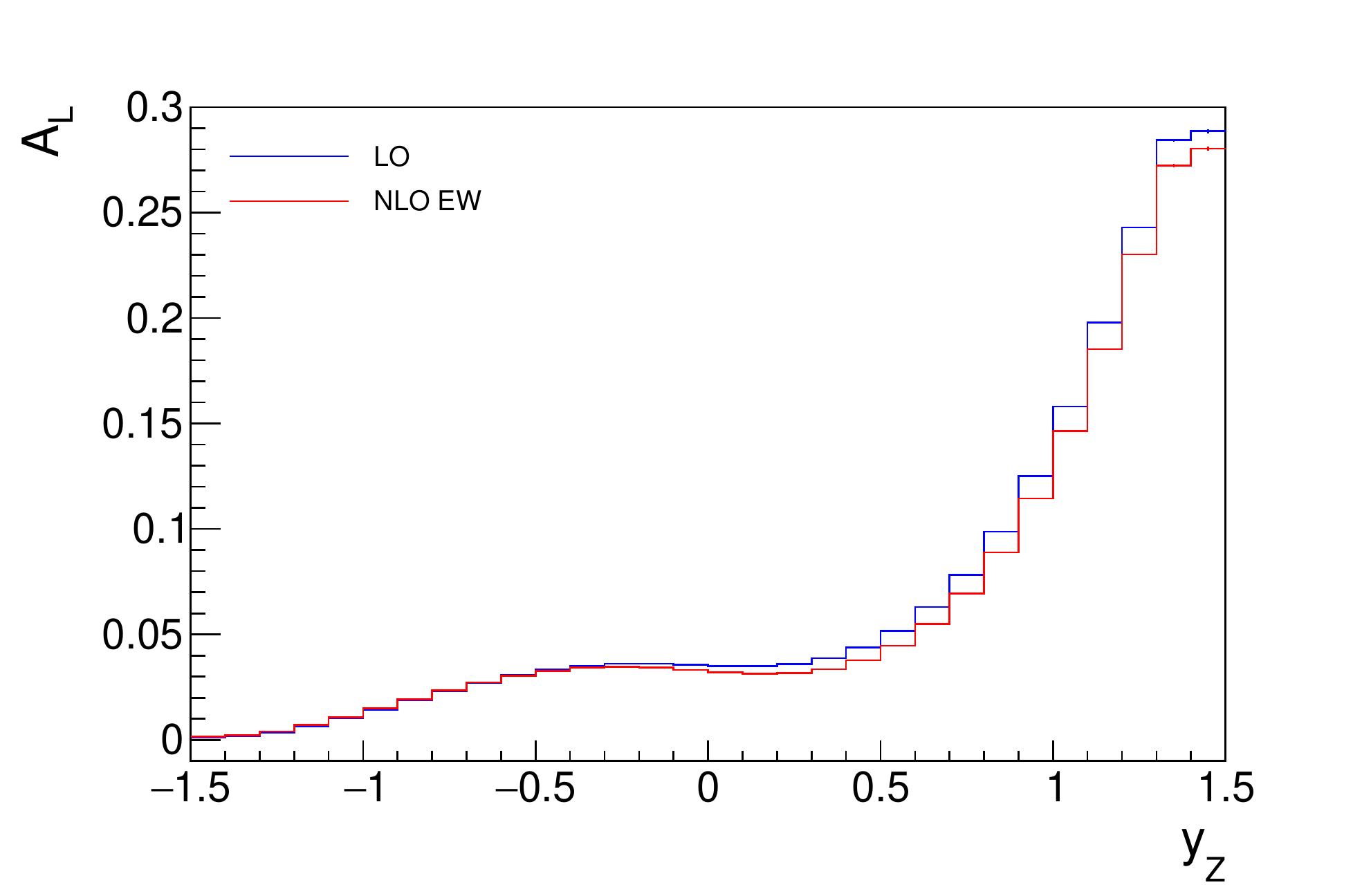} \hspace*{-7.5mm}
    & \hspace*{-7.5mm}\includegraphics[width=0.51\textwidth]{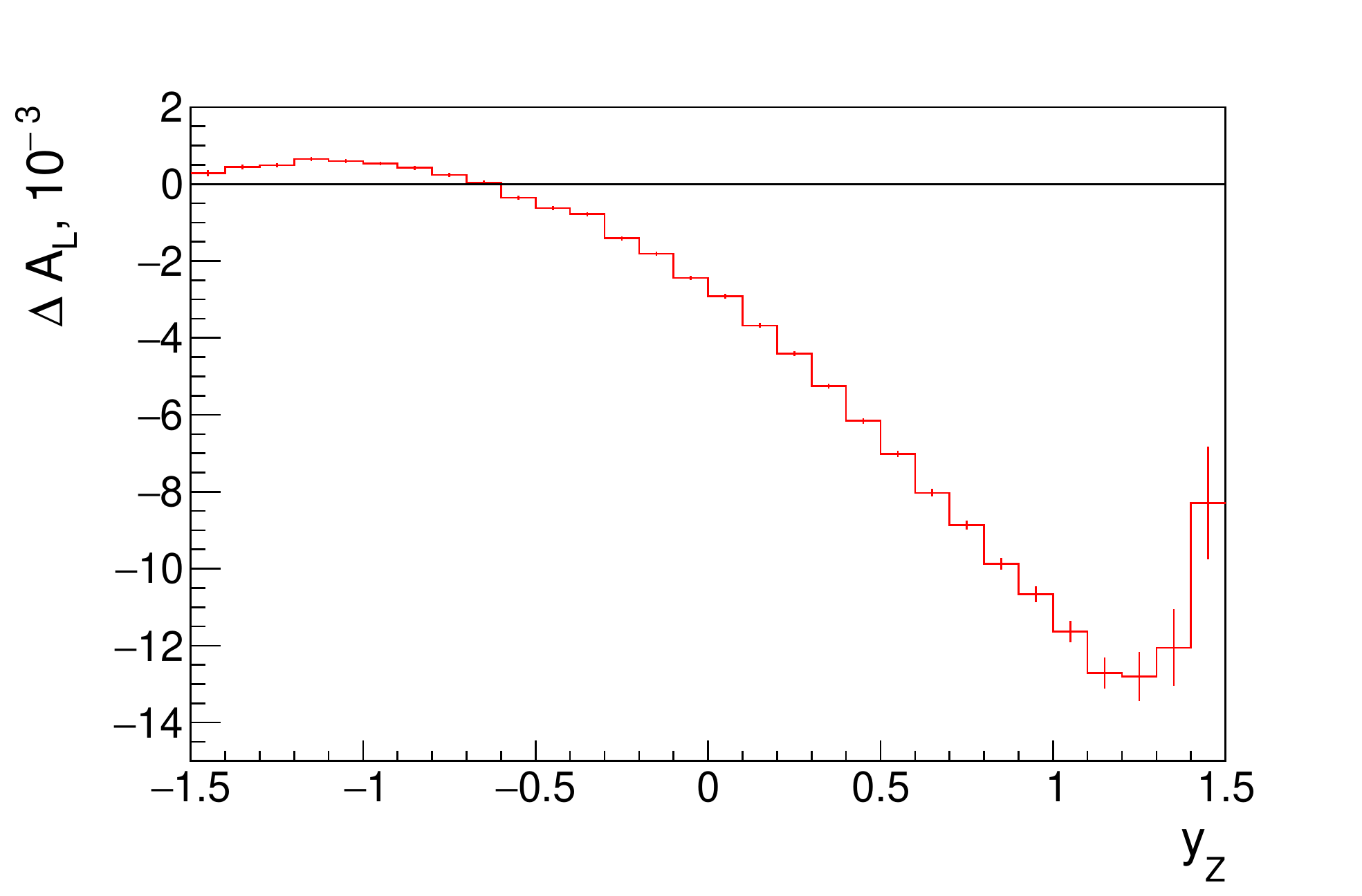}
\end{tabular}
\caption{ 
The $Z$ boson rapidity ${\mathrm y}_{\sss{\mathrm Z}}$
distribution for the single-spin asymmetry
${A}_{\mathrm{L}}({\mathrm y}_{\sss{\mathrm Z}})$   
    at the Born and NLO EW level (left panel)
    and corresponding difference  
    $\Delta {A}_{\mathrm{L}}({\mathrm y}_{\sss{\mathrm Z}}) $ (right panel).}
    \label{fig:Al-y34}
\end{figure*}

\begin{figure*}[!h]
\begin{tabular}{cc}
    \includegraphics[width=0.51\textwidth]{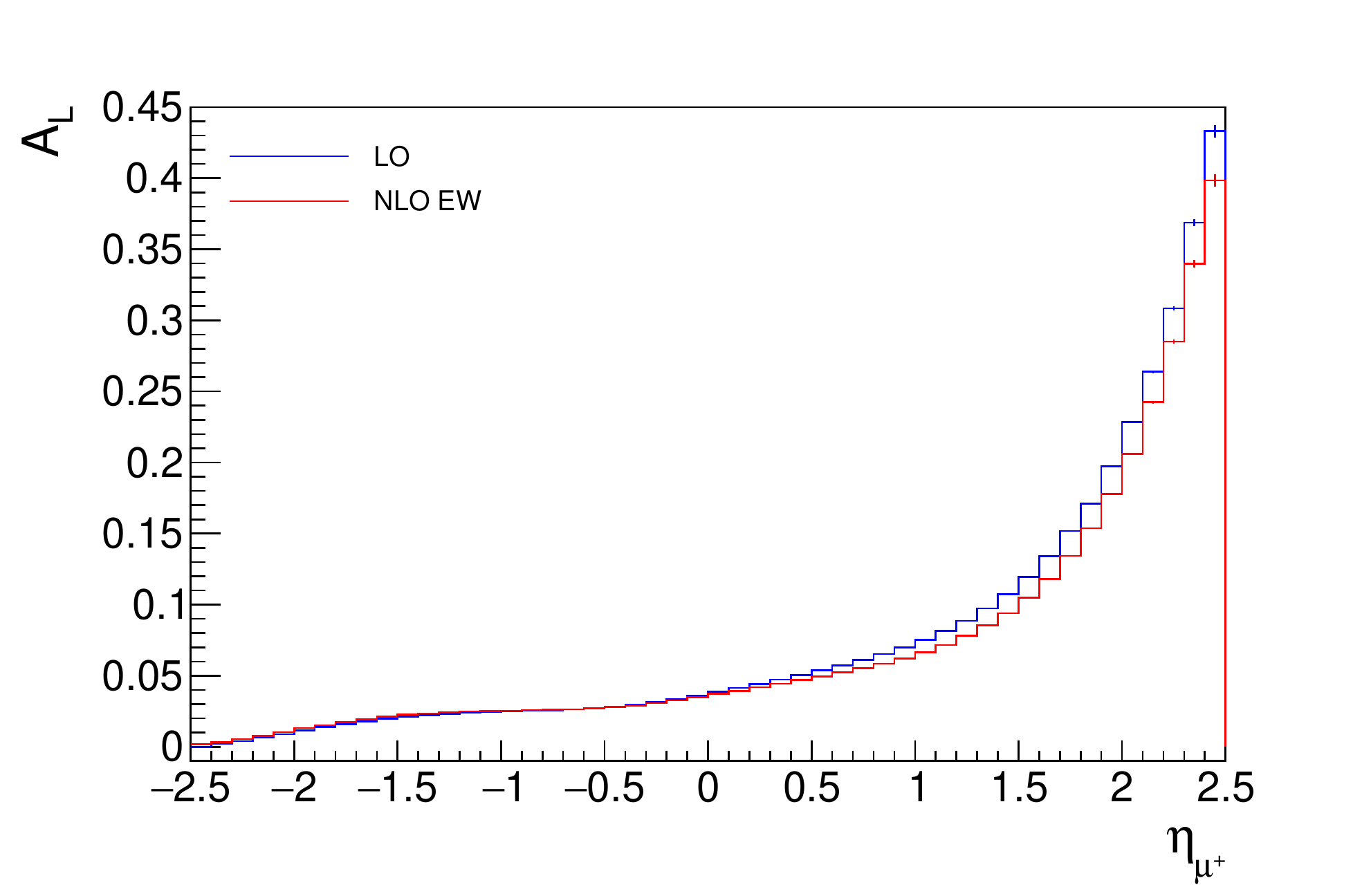} 
    \hspace*{-7.5mm}&\hspace*{-7.5mm}
    \includegraphics[width=0.51\textwidth]{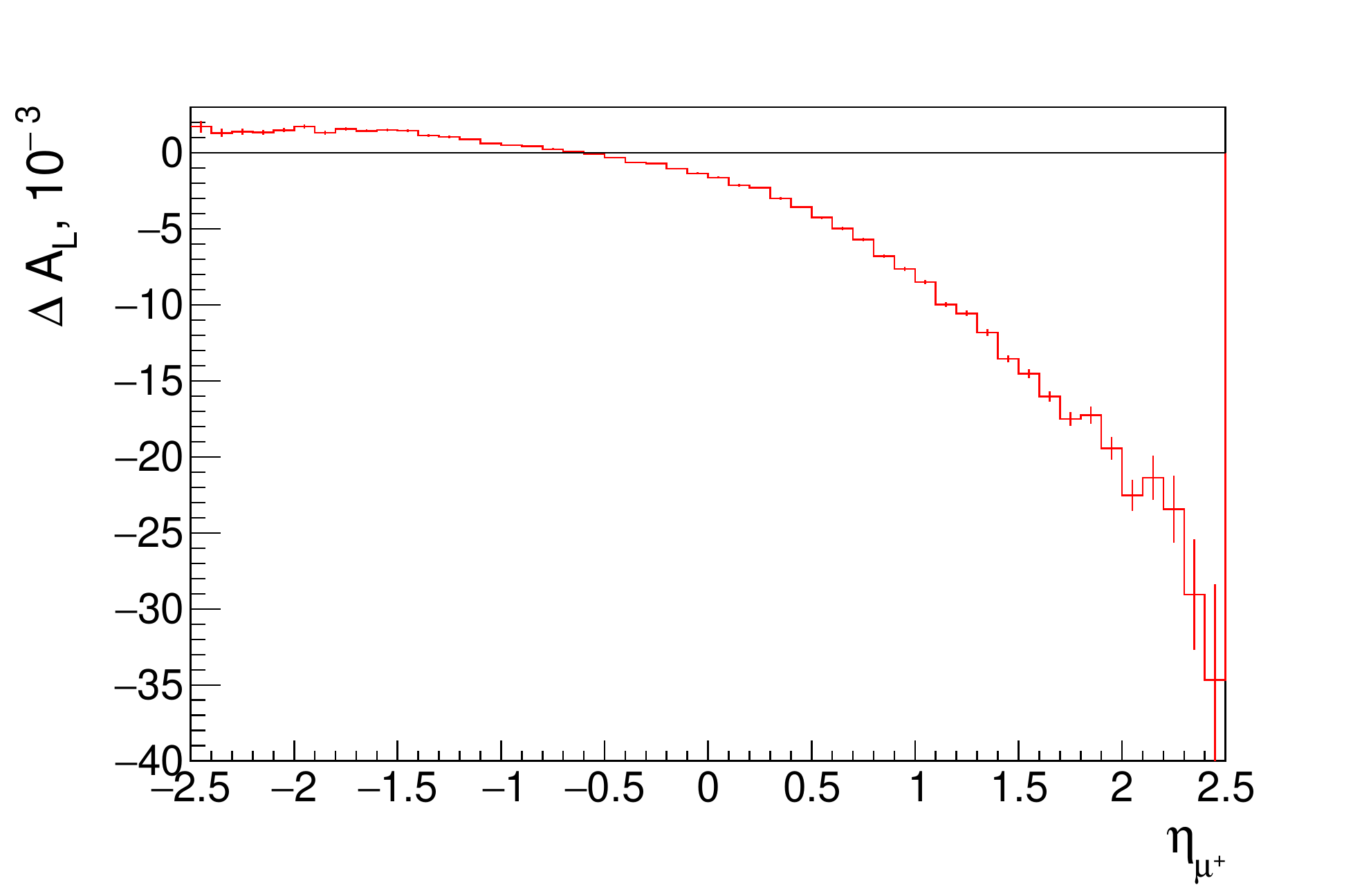}
\end{tabular}    
            \caption{
            The same as in Fig.\ref{fig:Al-y34} but for the anti-muon $\eta_{\mu^+}$ pseudorapidity.}
    \label{fig:Al-eta3}
\end{figure*}

\begin{figure*}[!h]
\begin{tabular}{cc}
    \includegraphics[width=0.51\textwidth]{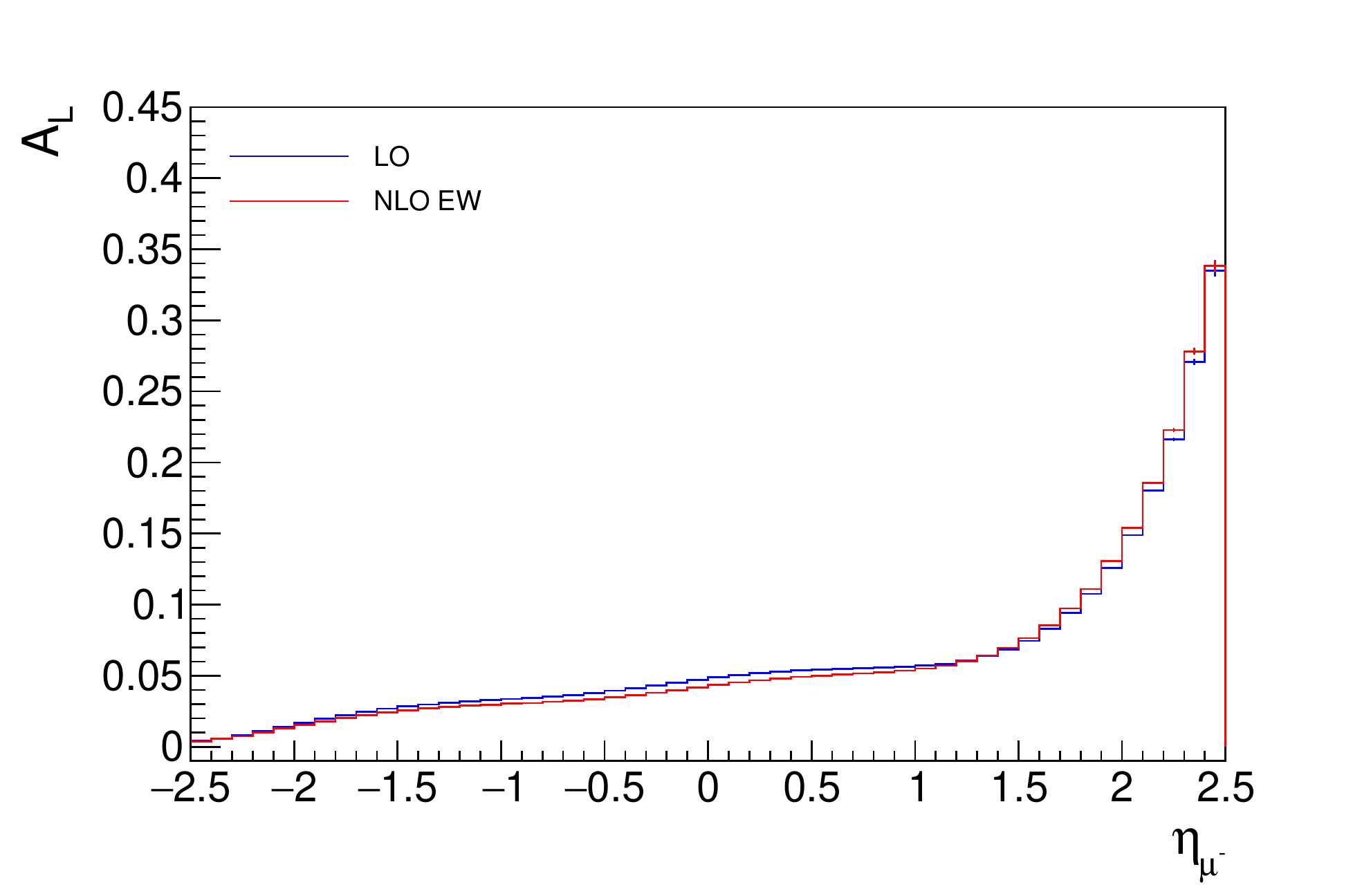}
        \hspace*{-7.5mm}&\hspace*{-7.5mm}
    \includegraphics[width=0.51\textwidth]{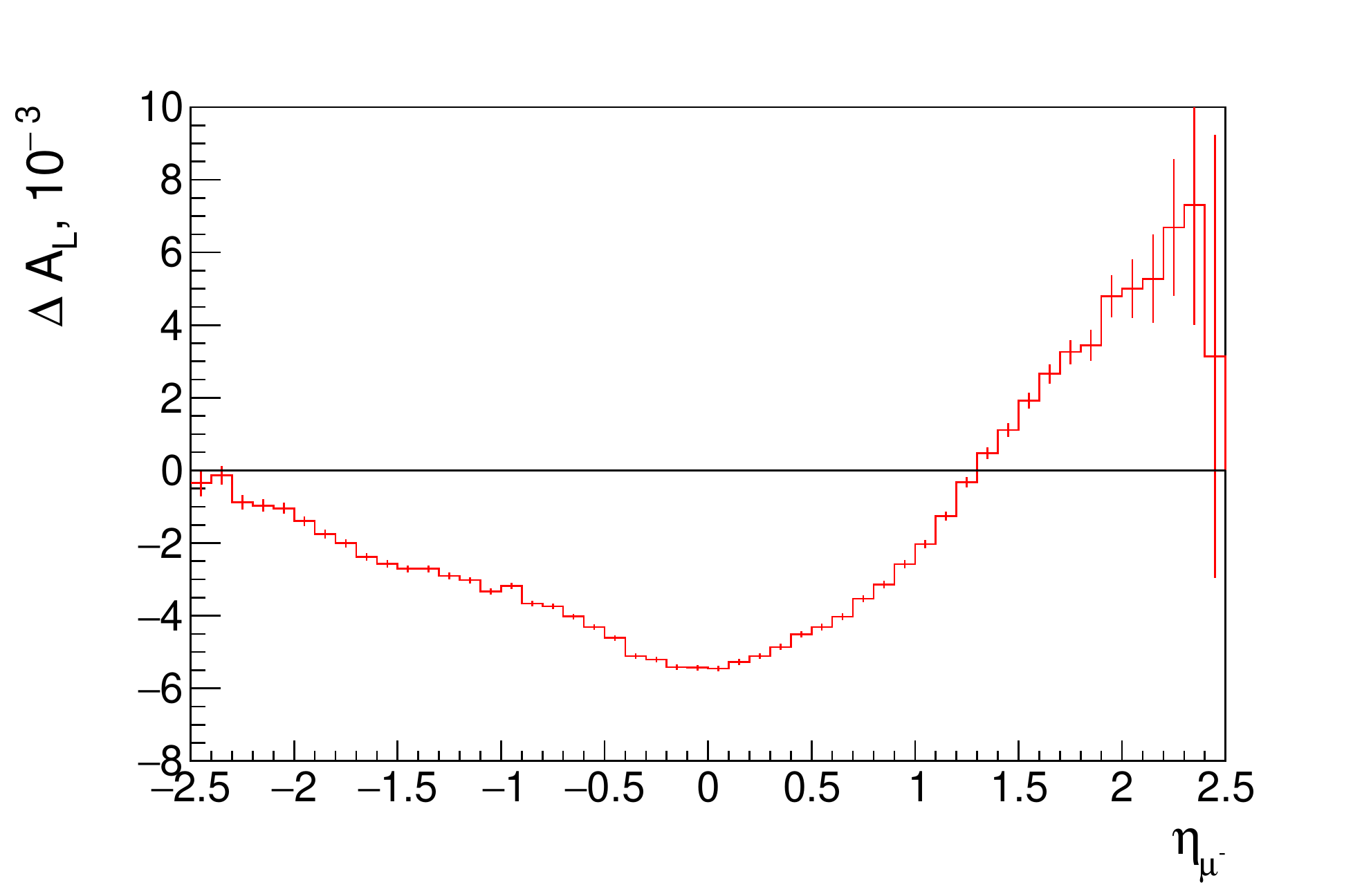}
\end{tabular}
    \caption{
The same as in Fig.\ref{fig:Al-y34} but for the muon $\eta_{\mu^-}$ pseudorapidity.
}
    \label{fig:Al-eta4}
\end{figure*}

\begin{figure*}[!h]
\begin{tabular}{cc}
    \includegraphics[width=0.51\textwidth]{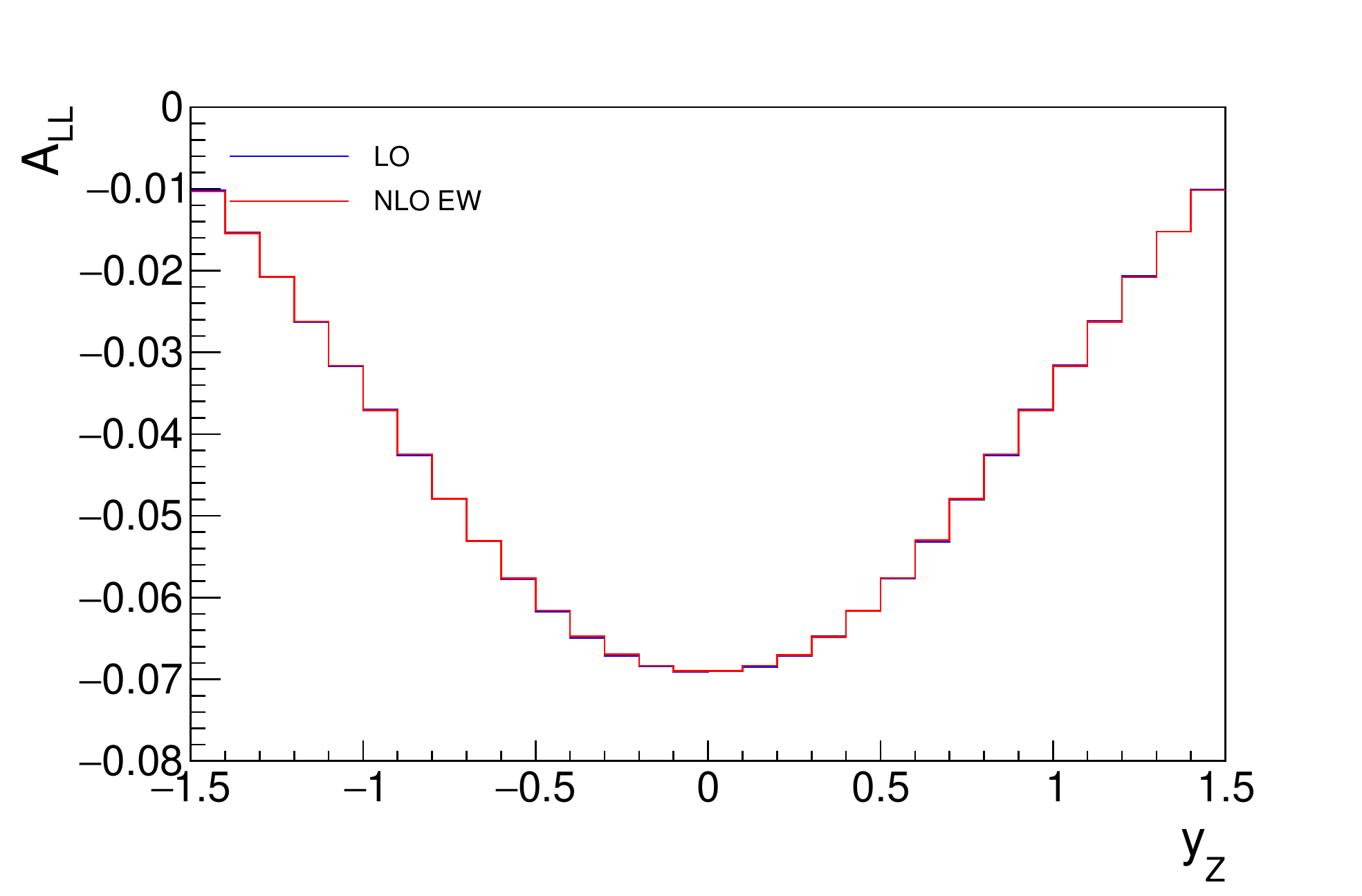}
    \hspace*{-7.5mm}&\hspace*{-7.5mm}
    \includegraphics[width=0.51\textwidth]{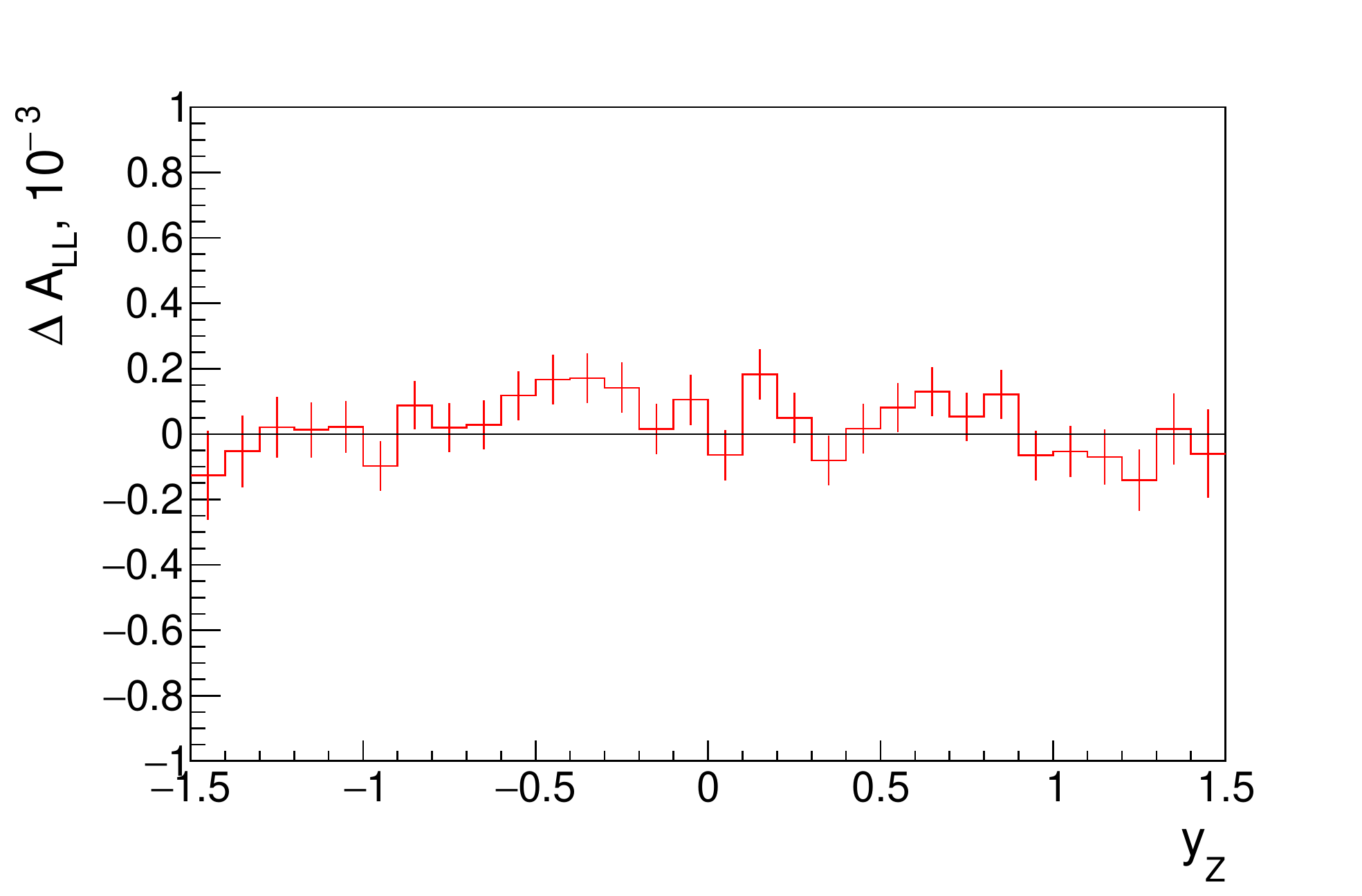}
\end{tabular}    
    \caption{
    The $Z$ boson rapidity ${\mathrm y}_{\sss{\mathrm Z}}$
distribution for the single-spin asymmetry
${A}_{\mathrm{LL}}({\mathrm y}_{\sss{\mathrm Z}})$   
    at the Born and NLO EW level (left panel)
    and corresponding difference  
    $\Delta {A}_{\mathrm{LL}}({\mathrm y}_{\sss{\mathrm Z}}) $ (right panel).  
}
    \label{fig:All-y34}
\end{figure*}

\begin{figure*}[!h]
\begin{tabular}{cc}
    \includegraphics[width=0.51\textwidth]{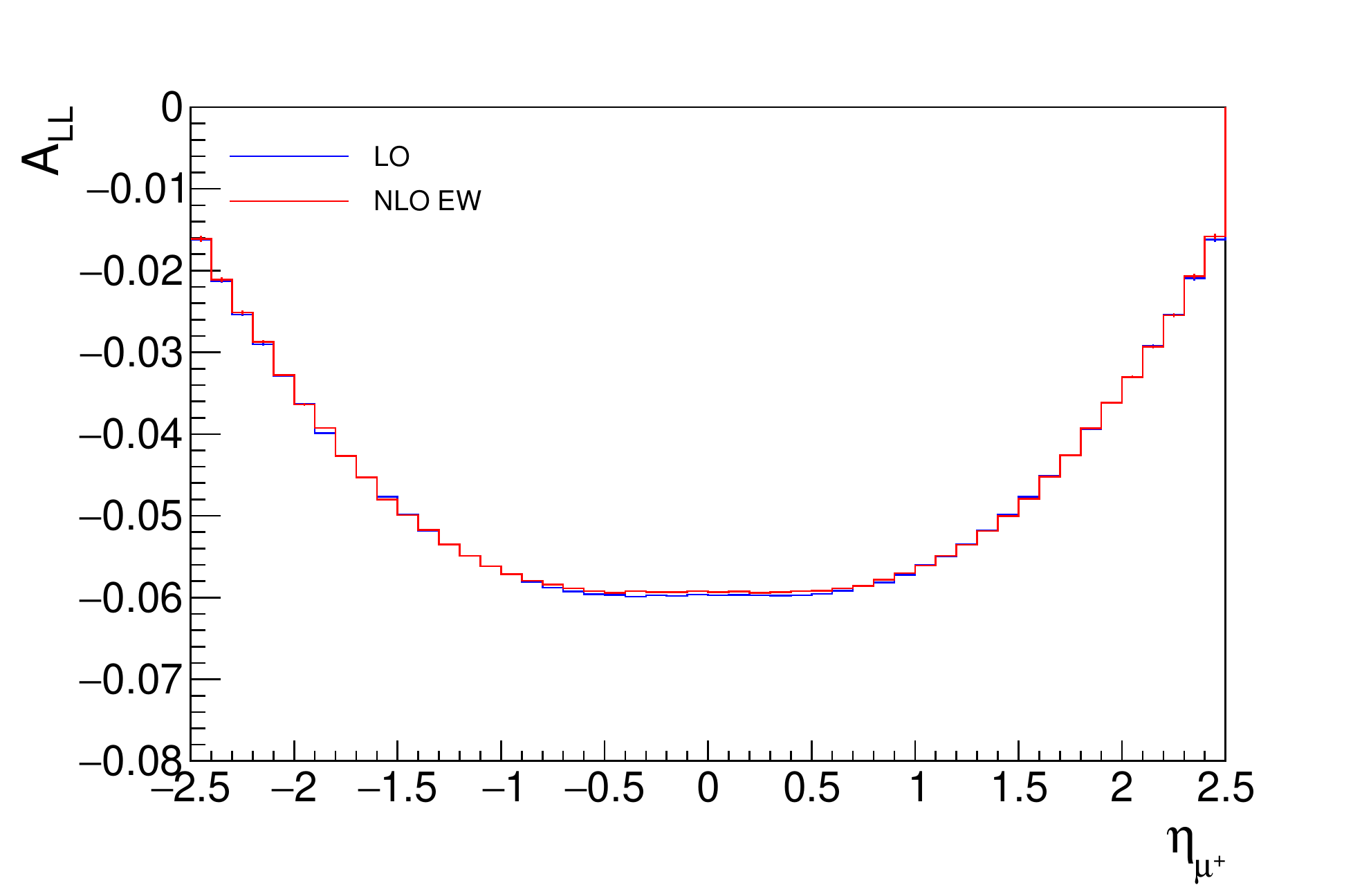}
 \hspace*{-7.5mm}&\hspace*{-7.5mm}
    \includegraphics[width=0.51\textwidth]{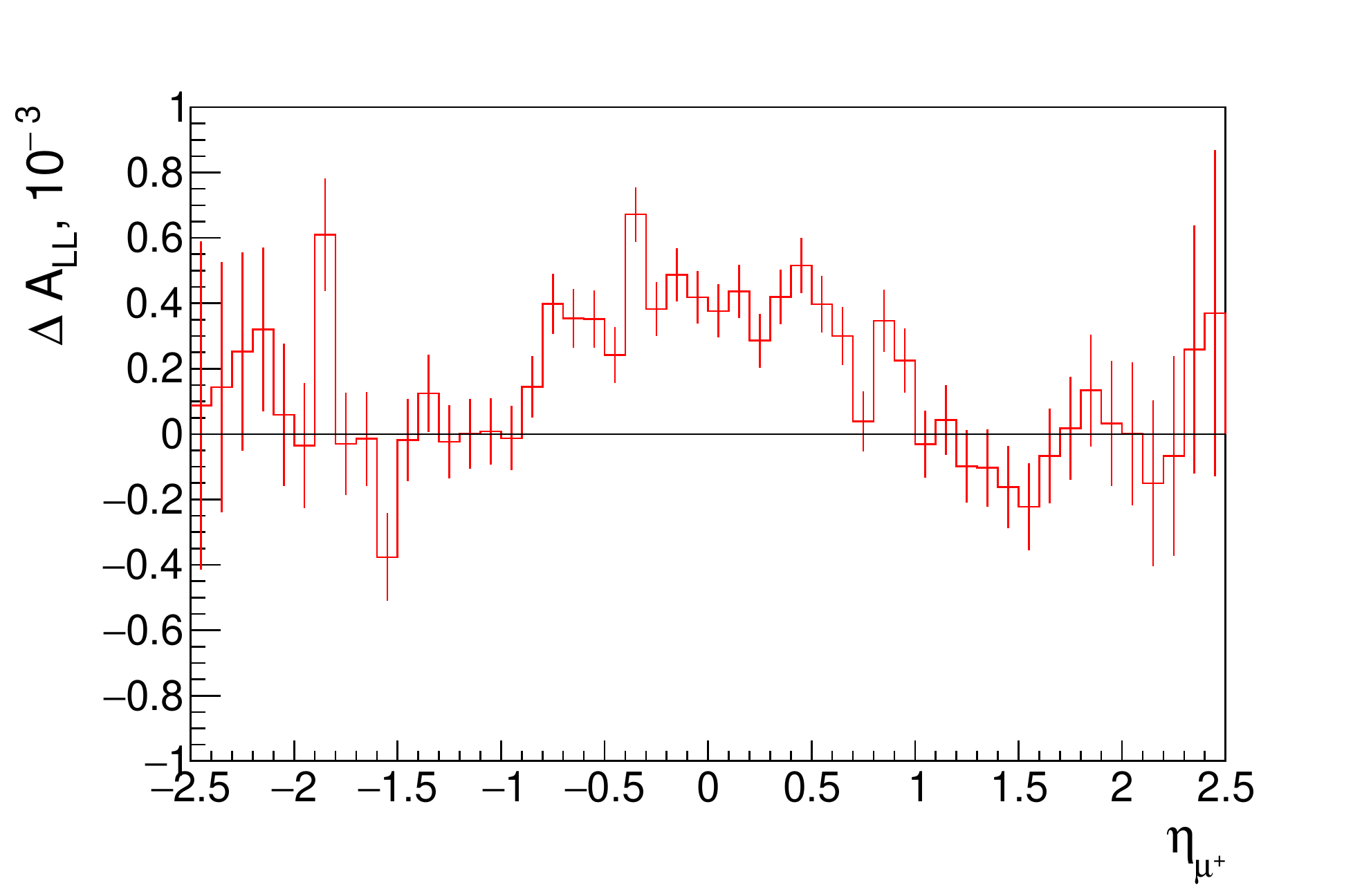}
\end{tabular}
\caption{
The same as in Fig.\ref{fig:All-y34} but for the anti-muon $\eta_{\mu^+}$ pseudorapidity.}
    \label{fig:All-eta3}
\end{figure*}

\begin{figure*}[!h]
\begin{tabular}{cc}
    \includegraphics[width=0.51\textwidth]{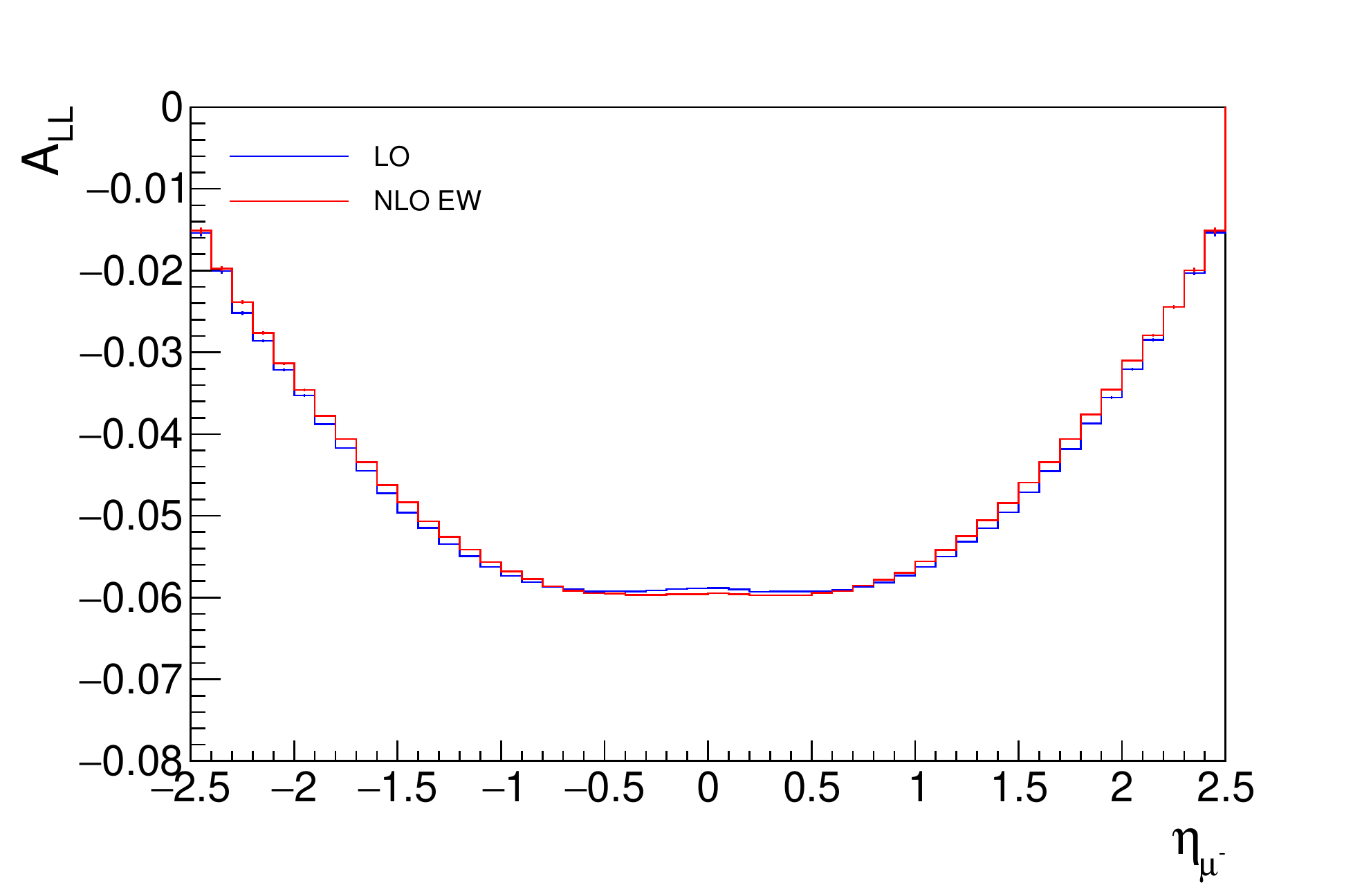}
 \hspace*{-7.5mm}&\hspace*{-7.5mm}
    \includegraphics[width=0.51\textwidth]{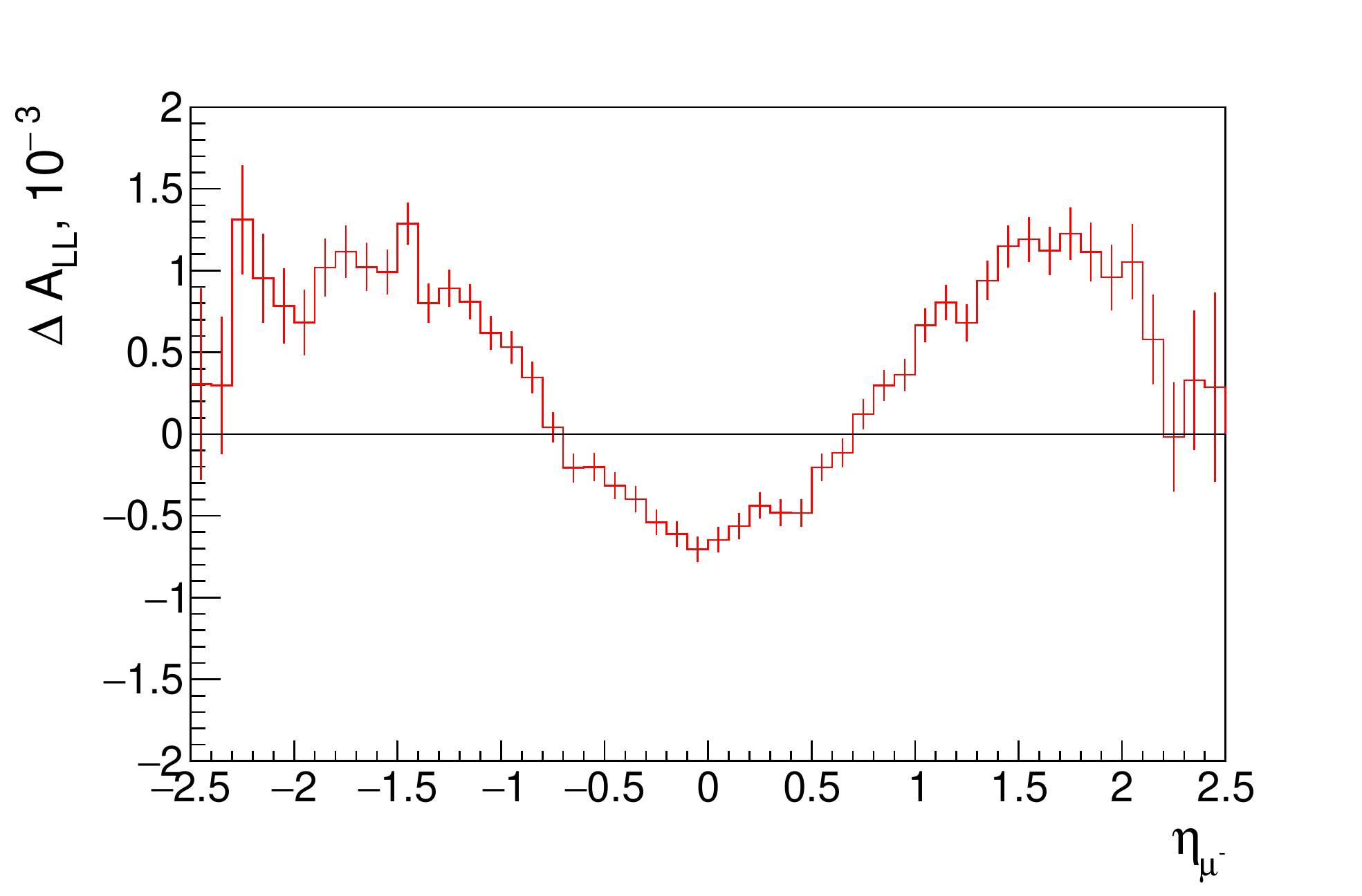}
\end{tabular}
\caption{
The same as in Fig.\ref{fig:All-y34} but for the muon $\eta_{\mu^-}$ pseudorapidity.
}
    \label{fig:All-eta4}
\end{figure*}

\begin{figure*}[!h]
    \centering
\begin{tabular}{cc}    
   \includegraphics[width=0.51\textwidth]{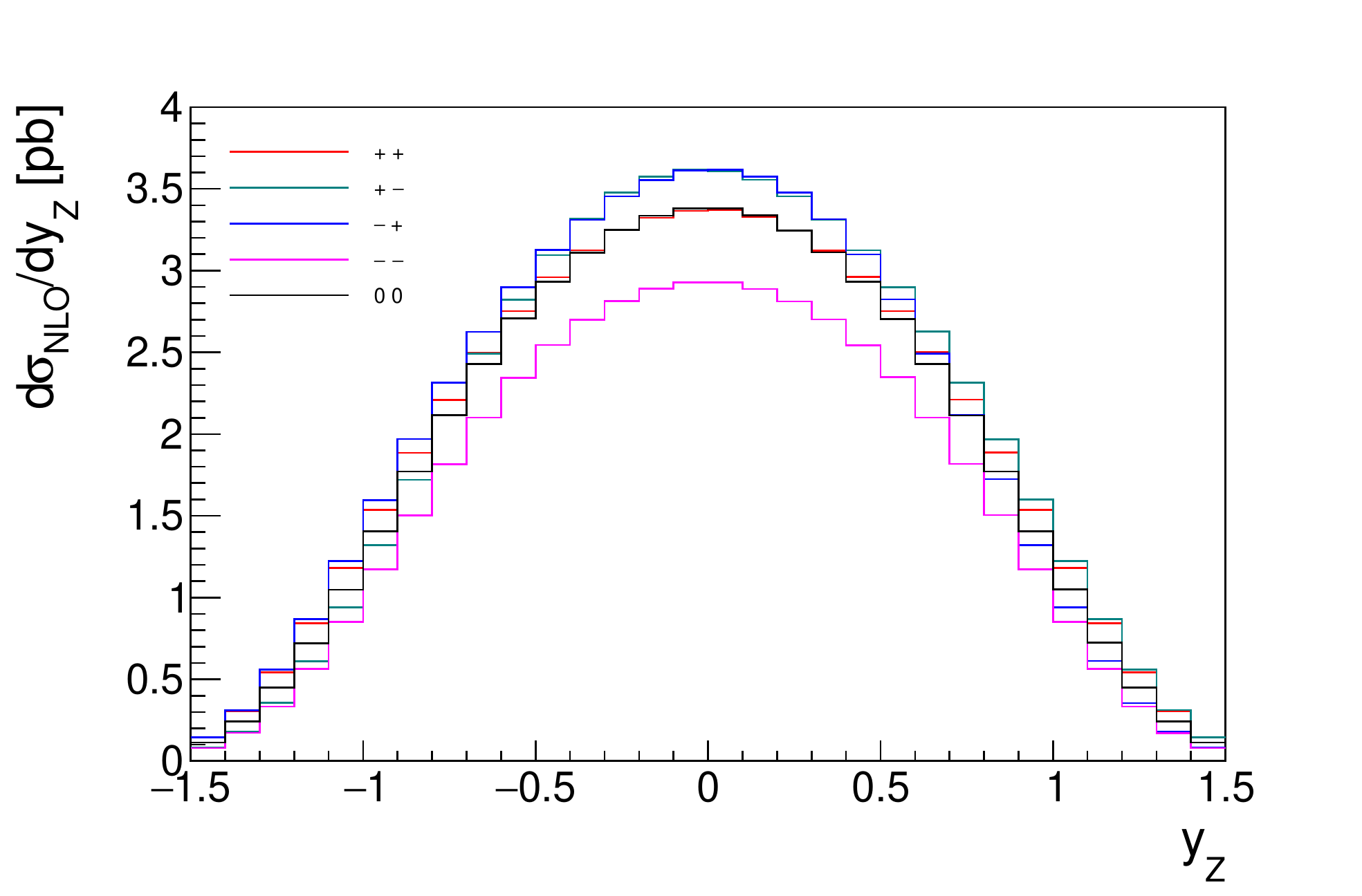}
       \hspace*{-7.5mm}&\hspace*{-7.5mm}
   \includegraphics[width=0.51\textwidth]{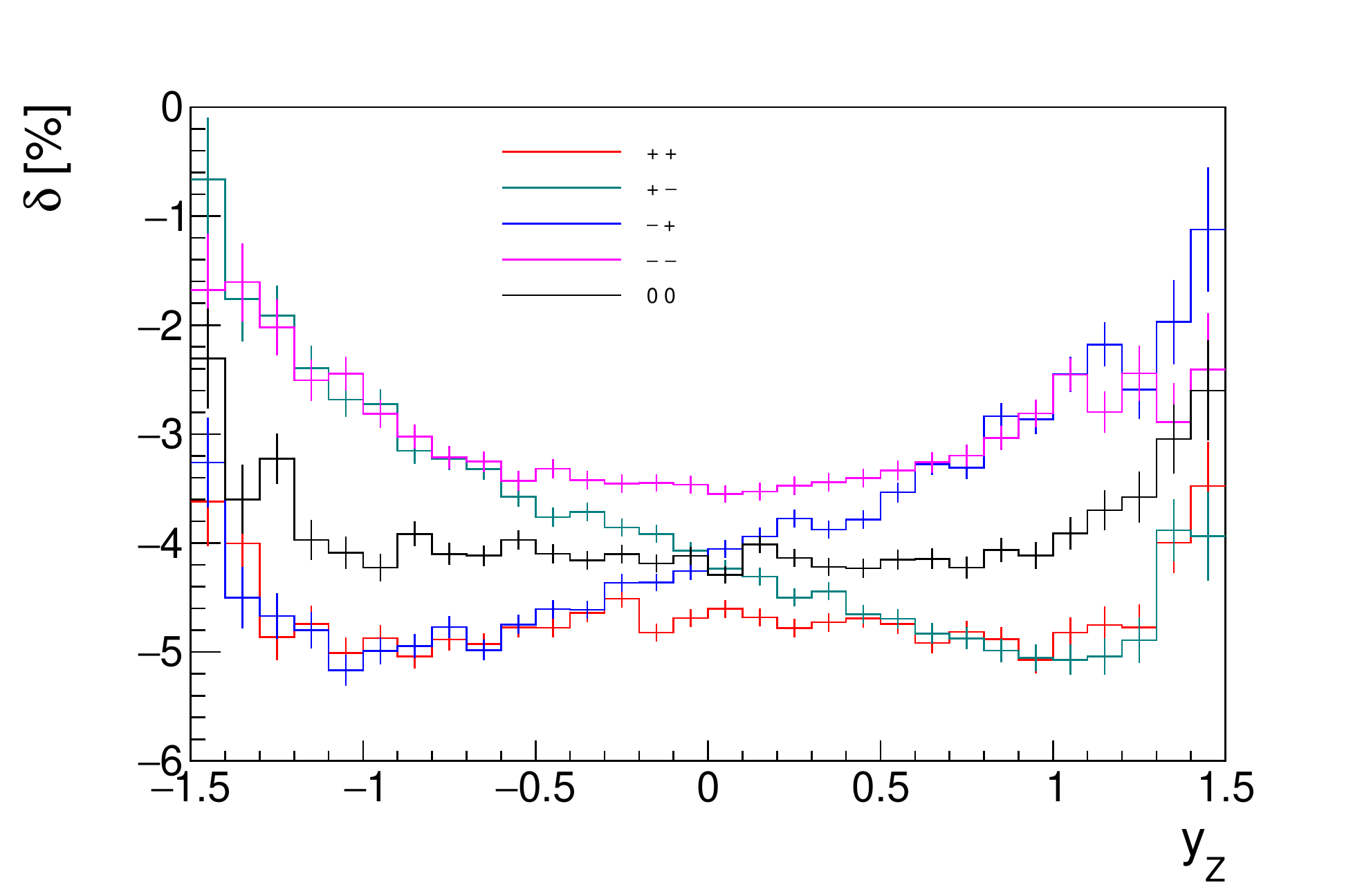}
\end{tabular}
            \caption{
    The $Z$ boson rapidity ${\mathrm y}_{\sss{\mathrm Z}}$
distribution for the  NLO EW cross section 
in pb (left panel)
    and for the relative corrections $\delta$ in $\%$ (right panel)
    for the components $(++),(+-),(-+),(--),(00)$.
}
    \label{fig:sigma-y34}
\end{figure*}

\begin{figure*}[!h]
\begin{tabular}{cc}  
    \includegraphics[width=0.51\textwidth]{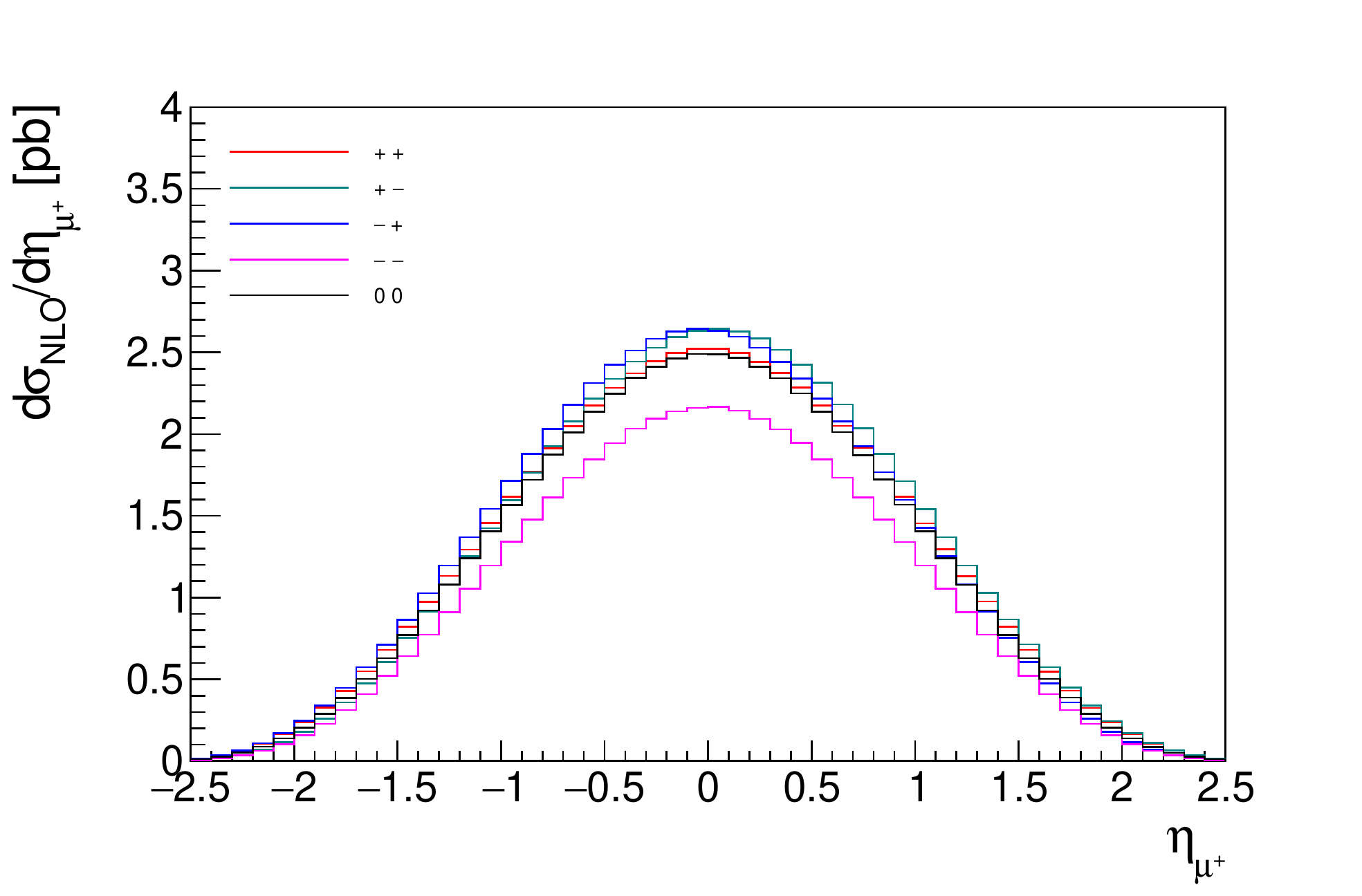}
        \hspace*{-7.5mm}&\hspace*{-7.5mm}
    \includegraphics[width=0.51\textwidth]{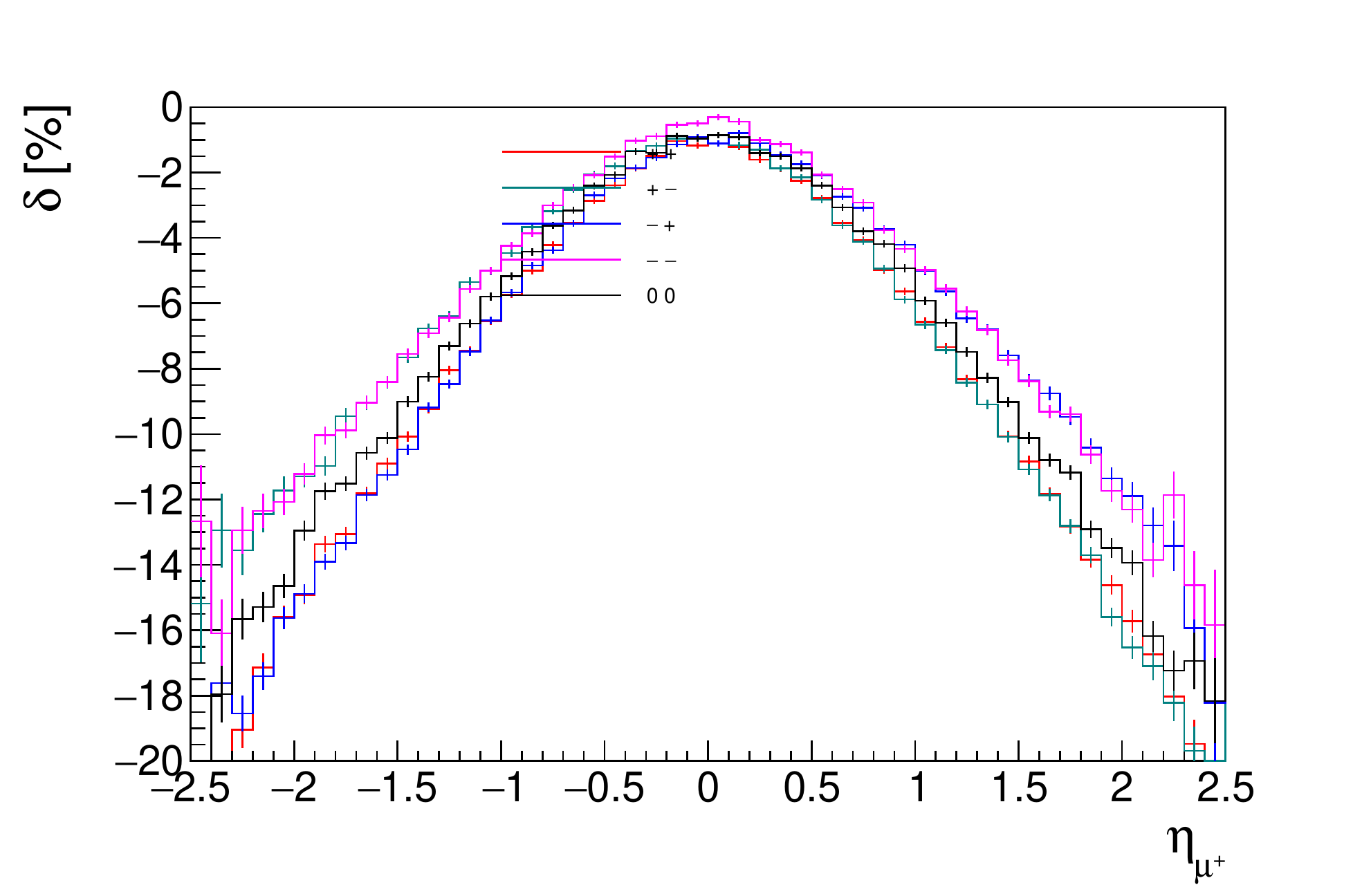}
\end{tabular}    
    \caption{The same as in Fig.\ref{fig:sigma-y34} but for the anti-muon $\eta_{\mu^+}$ pseudorapidity.}
    \label{fig:sigma-eta3}
\end{figure*}

\begin{figure*}[!h]
\begin{tabular}{cc}
    \includegraphics[width=0.51\textwidth]{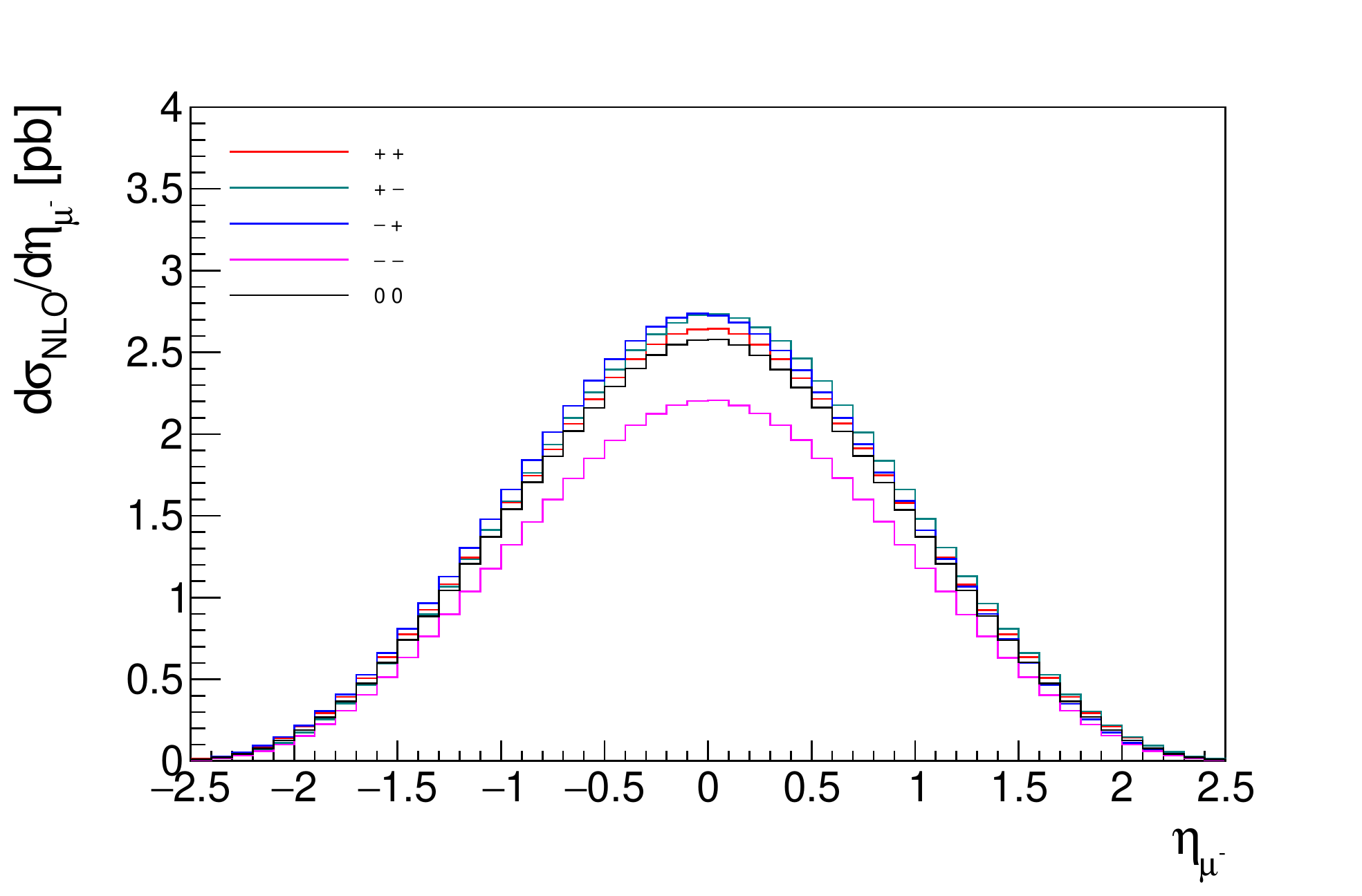}
        \hspace*{-7.5mm}&\hspace*{-7.5mm}
    \includegraphics[width=0.51\textwidth]{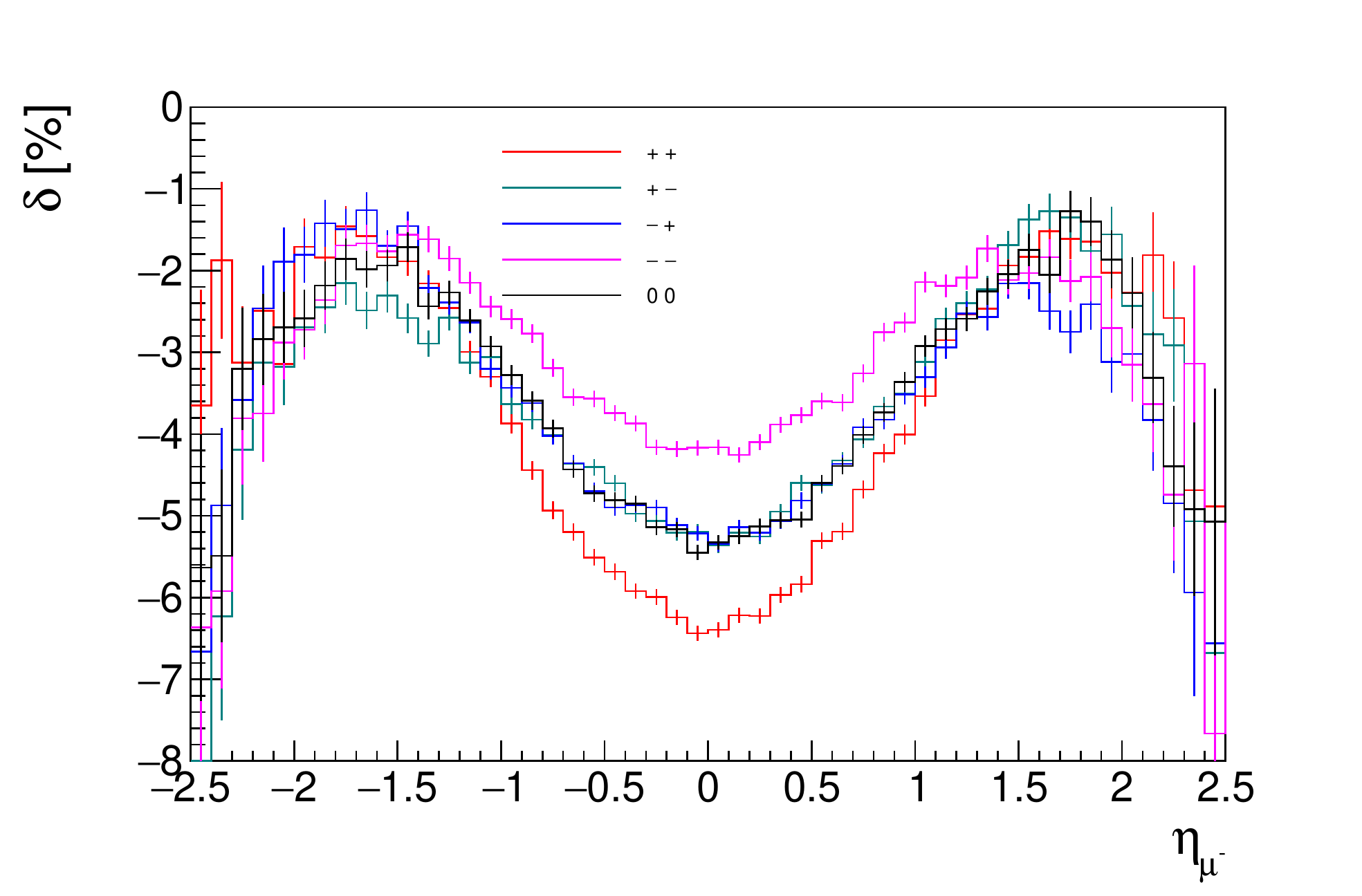}
\end{tabular}  
\caption{
The same as in Fig.\ref{fig:sigma-y34} but for the muon $\eta_{\mu^-}$ pseudorapidity.}
    \label{fig:sigma-eta4}
\end{figure*}

We demonstrate numerical calculations for the
$Z$ boson rapidity ${\mathrm y}_{\mathrm \sss Z}$ and lepton pseudorapidities $\eta_{\ell^\pm}$ distributions 
at the Born (LO) and NLO EW level
    and corresponding difference  
    $\Delta {A} = A^{\rm NLO~EW} - A^{\rm LO}$
for the single-spin asymmetry in Figs.~\ref{fig:Al-y34} -- \ref{fig:Al-eta4},
for the double-spin asymmetry in Figs.~\ref{fig:All-y34} -- \ref{fig:All-eta4}.
The same distributions  for cross sections in pb and corresponding relative corrections $\delta$ in $\% $  
are shown in Figs.~\ref{fig:sigma-y34} -- \ref{fig:sigma-eta4}.

A significant contribution of the NLO EW corrections to several distributions is observed.
Polarization asymmetries themselves do not change their signs -- ${A}_{\mathrm{L}}$ is always positive while ${A}_{\mathrm{LL}}$ is negative in whole kinematic region.

EW radiative corrections strongly depend on kinematic variables and
change the sign. 

Corrections to  ${A}_{\mathrm{LL}}$  for   rapidity $y_Z$ distribution
are compatible with zero
while for distributions  of pseudorapidities 
  $\eta_{\mu^+}$ and  $\eta_{\mu^-}$
the corrections are mostly  positive and oscillating near mean value of about 1\%.

The partial differential cross-sections as a function of rapidity $y_Z$, pseudorapidities $\eta_{\mu^+}$ and  $\eta_{\mu^-}$ 
are sensitive to the polarization of incoming particles. 
The line $00$ shows unpolarized case while other lines are for the 100\% polarized beams. 
One sees that relative corrections $\delta$ are negative and strongly depend on beam polarization,
and vary from $-3\%$ to $-5\%$  in the central region of variable $y_Z$ and from $-12\%$ to $-20\%$ 
in the forward region of pseudorapidity  $\eta_{\mu^+}$.
Radiative corrections are symmetric for $\eta_{\mu^+}$, $\eta_{\mu^-}$ and unsymmetric for $y_Z$.

\section{Conclusion}
\label{sect-concl}
In the paper for the first time
the study of spin effects at NLO EW level
in neutral current Drell-Yan processes
in collisions of longitudinally polarized hadrons 
is presented.
We have shown numerical results for observables obtained by MC event generator {\ReneSANCe}. 
The effects of complete one-loop electroweak radiative corrections to NC DY processes 
are significant.
 
Obtained NLO EW corrections can be used for reduction of the systematic uncertainty in measurement of polarized   parton distributions.

We also expect a valuable effect of EW radiative corrections in polarized production of charged vector boson and plan to study it.
Another direction of our investigation is to include effects from transverse polarization, which are strongly related to transverse-momentum-dependent distribution of partons.

\section{Funding}
\label{sec:funding}
The research is supported by the Russian Science Foundation (project No. 22-12-00021). 
 
\section{Acknowledgements}
\label{sec:acknowledgements}
We are grateful to A.~Arbuzov for discussion of the physical results.

\providecommand{\href}[2]{#2}
\end{document}